\documentstyle[12pt]{article}

\newcommand{\lsim}{\mathrel{\lower4pt\hbox{$\sim$}}
\hskip-12.5pt\raise1.6pt\hbox{$<$}\;}

\newcommand{\gsim}{\mathrel{\lower4pt\hbox{$\sim$}}
\hskip-12.5pt\raise1.6pt\hbox{$>$}\;}

\begin{document}
\baselineskip 18pt plus 2pt

\noindent \hspace*{10cm}UCRHEP-T185\\

\begin{center}
{\bf Probing the Flavor-Changing $tc$ vertex via Tree-Level Processes:
$e^+e^-\to t\bar c\nu_e\bar\nu_e$, $t\bar ce^+e^-$ and $t\to cW^+W^-$ } 

\bigskip\bigskip

S. Bar-Shalom$^a$, G. Eilam$^b$, A. Soni$^c$ and J. Wudka$^a$


$^a$ Physics Dept., University of California, Riverside CA 92521.\\
$^b$ Physics Dept., Technion-Israel Inst.\ of Tech., Haifa 32000, Israel.\\
$^c$ Physics Dept., Brookhaven Nat.\ Lab., Upton NY 11973, USA.
\date{\today}

\end{center}

\bigskip\bigskip

\abstract{
\begin{quote}
{\bf Abstract}: The reactions $e^+e^-\to t\bar c\nu_e\bar\nu_e$, $t\bar
ce^+e^-$ are very sensitive probes of the flavor-changing-scalar
couplings which can occur in a model with one extra Higgs doublet. At
the Next Linear Collider, with a center of mass energy of
$\sqrt{s}=0.5$--2 TeV, several hundreds and up to thousands of such 
events may be produced
if the mass of the light neutral Higgs is a few hundred GeV\null. We
also briefly comment on the decays $t\to cW^+W^-$, $t\to cZZ$. All of
these reactions are severely suppressed in the Standard Model.
\end{quote}
}


\newpage

Understanding the nature of the scalar sector in electroweak theories
and searching for flavor-changing (FC) currents are clearly important goals
of the next generation of high energy colliders. The purpose of this
work is to point out that the reactions:
\begin{equation}
e^++e^- \to t\bar c\nu_e\bar\nu_e;~\bar tc\nu_e\bar\nu_e;~t\bar
ce^+e^-;~\bar tce^+e^- \label{eppem}
\end{equation}
are extremely sensitive probes for such investigations and should be accessible to the next generation of $e^+$-$e^-$ linear
colliders (NLC) currently being envisaged \cite{nlc}.

As is well known, though there are stringent experimental constraints against
the existence of tree level flavor-changing-scalar (FCS) transitions
involving the light quarks \cite{weinb,sher1,soni1}, analogous constraints
involving the top quark are essentially non-existent.
In fact, it is natural to imagine that FCS
interactions are proportional to the masses of the fermions
participating at the vertex \cite{sher1}; in such a scenario the large top
mass makes it much more susceptible to
FC transitions. This reasoning has led various authors to stress the
importance of searching for tree-level FCS
interactions involving the top-quark, especially the top-charm ones
\cite{savage,houlin}. Our study indicates that experimental
investigations of the reactions in (\ref{eppem}) could be very
useful in this regard.

A mild extension of the Standard Model (SM) in which one extra
scalar doublet is added, allows for large, tree-level FCS 
interactions \cite{luke}. Therefore, the
two Higgs doublet model (2HDM) scalar potential is usually 
constrained by a discrete
symmetry \cite{weinb} whose only role is to forbid tree-level 
flavor-changing-scalar-currents. If one does not impose such a discrete symmetry by hand, one arrives
at a version of the 2HDM, called Model~III, wherein the up-type and the
down-type quarks are allowed simultaneously to couple to more than one
scalar doublet \cite{luke}. The diagonalization of the quark mass
matrices does not automatically ensure the diagonalization of the
couplings with each single scalar doublet. Both up and down type quarks
may then have FC couplings and the corresponding
Yukawa Lagrangian in this model is \cite{luke,soni1}
\begin{equation}
{\cal L}_Y^{FC}=\xi_{ij}^U {\bar Q}_{i,L} {\tilde \phi}_2 U_{j,R} +    
\xi_{ij}^D {\bar Q}_{i,L} \phi_2 D_{j,R} + h.c. ~, 
\end{equation}
where $\phi_2$ denotes the second scalar doublet, ${\tilde \phi}_2 
\equiv i \tau_2 \phi_2$, $Q$ stands for the quark doublets, and $U$ 
and $D$ for charge 2/3 and (-1/3) quarks singlets; $i,j=1,2,3$ 
are the generation indices and $\xi$ are $3\times 3$
matrices parameterizing the strength of FC neutral scalar vertices. 
The spectrum of the scalar sector in this model consists of two 
neutral Higgs scalars, denoted as $h,H$ and a pseudoscalar $A$. 
In addition, the model has two charged scalars $H^{\pm}$.   

The experimental constraints can be accommodated simply by imposing a
hierarchy among the FC vertices \cite{sher1} whose strength is 
correlated to the masses of the participating quarks. We will thus take
\begin{equation}
\xi_{ij}^{U,D}=g_W \left({\sqrt {m_im_j}}/m_W \right) \lambda_{ij} \label{csa} ~. 
\end{equation}
which is often called the Cheng-Sher Ansatz (CSA) \cite{sher1,soni1,luke}.

In this scenario all our ignorance regarding the FCS vertices is 
in the couplings $\lambda_{ij}$ which are free parameters
to be experimentally determined. 
Assuming for now that they are real, there are six such
couplings: $\lambda_{sd}$, $\lambda_{bs}$, $\lambda_{bd}$, $\lambda_{cu}$,
$\lambda_{tu}$ and $\lambda_{tc}$. Detailed examination of low 
energy experimental data,
primarily from $\Delta$flavor${}=2$ processes, leads to $\lambda_{sd}$, $\lambda_{bs}$, $\lambda_{cu} \lsim 0.1$
\cite{soni1}. Existing experimental information does not provide
any useful constraints on $\lambda_{tc}$; in particular, we may well have
$\lambda_{tc}\sim {\cal O}(1)$ \cite{lambdatu}.
In this Letter we will show that if $\lambda_{tc}\sim {\cal O}(1)$,
experiments on reaction (\ref{eppem}) at the NLC can lead to
spectacular signatures.

Our study shows that an extremely interesting feature of the reactions 
in (\ref{eppem}) is that (within Model~III) the cross-section for these 
reactions can be much larger than the simple s-channel reaction $e^+e^- 
\to t \bar c$ (see Atwood {\it et. al.} in 
\cite{savage}). Indeed $\sigma^{\nu \nu tc} \equiv \sigma(e^+e^- \to t 
\bar c \nu_e {\bar {\nu}}_e + \bar t c \nu_e {\bar {\nu}}_e)$ is about 
two orders of magnitude larger than $\sigma(e^+e^- \to t \bar c + \bar 
t c)$ over a large region of parameter space
 of Model~III; also $\sigma^{eetc} \equiv \sigma(e^+e^- \to t \bar c e^+ e^- 
+ \bar t c e^+ e^-)$ is about one order of magnitude bigger than 
$\sigma(e^+e^- \to t \bar c + \bar t c)$. Moreover, while the 
cross-section for producing  
$t \bar c$ drops as $\sqrt s$ increases, the cross-sections 
$\sigma^{eetc}$ and $\sigma^{\nu \nu t c }$ increase with energy in 
the range $0.5 ~{\rm TeV}<\sqrt s<2 ~{\rm TeV}$. Thus, even if no 
$t \bar c$ events are detected at 
$\sqrt s =0.5$ TeV via $e^+e^- \to t \bar c$, there is still a 
strong motivation to look for
signatures of (\ref{eppem}), especially at somewhat higher energies. 

In exploring the reactions $e^+e^-\to t\bar c\nu_e\bar\nu_e$, $t\bar c
e^+e^-$ we will use the effective vector boson approximation (EVBA)
\cite{cahn}. The salient features of reaction
(\ref{eppem}) are then well approximated by the simpler fusion
reactions
\begin{equation}
W^+W^-, Z Z \to t \bar c, \bar tc
\end{equation}
The corresponding cross sections for the reactions in
(\ref{eppem}) are then calculated by folding in the distribution
functions $f^V_{h_V}$, for a vector boson $V$ ($W$ or $Z$) with helicity $h_V$ \cite{footvv}.

The EVBA has been extensively studied in the production of a $t\bar t$
pair \cite{wwtt}. There is, however,
a significant difference between fusion leading to $t\bar c$ and to 
$t\bar t$ primarily due to the appreciable
difference in the threshold of the two-reactions, due to $m_t\gg
m_c$. For $t\bar c$ the fraction of the incoming
vector-boson energy, $x=\sqrt{\hat s/s}$ ( $\hat s$ is the
c.m.\ energy squared in the $VV$ c.m.\ frame and $s$ in the $e^+e^-$
c.m.\ frame), can drop below 0.05 near threshold, for $\sqrt{s}\gsim 0.8$
TeV\null. In this small-$x$ range ($x\lsim0.05$) the distribution
functions are overestimated within the leading log approximation
\cite{wwtt,wapprox1}. We will therefore use the distribution functions
which retain higher orders in $m^2_V/s$ as given in
Ref.~\cite{wapprox1}. In the cross section $ \sigma( VV \to t \bar t) $ 
the dominant contribution $\propto (m_t/m_V)^4$ is generated
by the longitudinal vector-boson contributions for which the
polarization vector can be approximated by $\epsilon^\mu_0(k)
\simeq k^\mu/m_V $. This approximation does
not necessarily hold for the reaction $VV\to t\bar c$ for which
$m^2_V/\hat s \approx m^2_V/m^2_t$ near threshold. In particular, 
the cross-section for the reaction $VV\to {\cal
H}\to t\bar c$ (${\cal H}$ denotes a neutral Higgs particle) scales
like $|\epsilon^{V^1}_{h_{V^1}} \cdot \epsilon^{V^2}_{h_{V_2}}|^2$
(see below).
Thus not only is the $(m_t/m_V)^4$ factor absent, but also, near
threshold, the contribution from the transversely polarized $V$'s is
comparable to that of the longitudinal $V$'s. We will therefore perform
an exact calculation of $\hat\sigma(VV\to {\cal H}\to t\bar c)$ keeping
all possible polarizations of the two colliding vector-bosons.

It is interesting to note that while at tree-level, $\sigma^{eetc}=0$ in the 
SM, the parton level reaction $W^+W^- \to t \bar c$ can proceed at tree-level, 
via Fig.~1a. 
However, numerically, due to GIM suppression, the results is found to be too small to be of 
experimental relevance: $\sigma^{\nu \nu tc}_{\rm SM} \approx 10^{-5}-10^{-4}$ fb 
for $\sqrt s = 0.5 - 2$ TeV \cite{longerver}. We will henceforth neglect the SM contribution.

In Model~III there is an additional important tree-level contribution
(see Fig.~1b), originating from $VV\to {\cal H}\to t\bar c$, $\bar
tc$. Choosing for simplicity
$\lambda_{tc}=\lambda_{ct}=\lambda_R +
i\lambda_I$, the relevant terms of the Model~III Lagrangian with the CSA become:
\begin{eqnarray}
{\cal L}_{{\cal H}tc}&=&-\frac{g_W}{\sqrt 2} \frac{{\sqrt {m_t m_c}}}{m_W}
f_{\cal H} \; {\cal H} \; \bar t (\lambda_R+i\lambda_I \gamma_5) c
\label{htc}~,\\ 
{\cal L}_{{\cal H} VV}&=&-g_W m_W C_V c_{\cal H} \; {\cal H} g_{\mu\nu} V^{\mu}
V^{\nu} \label{hww}~, 
\end{eqnarray}

\noindent where $C_{W;Z}=1;m_Z^2/m_W^2$, $f_{h;H}\equiv \cos{\tilde
{\alpha}}; 
\sin{\tilde {\alpha}}$ and $c_{h;H} \equiv \sin{\tilde {\alpha}} 
;-\cos{\tilde {\alpha}}$. The mixing angle ${\tilde {\alpha}}$
is determined by the Higgs potential. Note that the
pseudoscalar $A$ does not couple to gauge bosons and is therefore
irrelevant for the reactions at hand.

Within Model~III $\sigma^{\nu \nu tc},
\sigma^{eetc}({\tilde {\alpha}} \to 0~or~\pi/2) \to 0$ at the tree-level. 
For definiteness, we will present our numerical results for 
${\tilde {\alpha}}=\pi/4$ \cite{longerver,footalpha}. In calculating the 
cross sections we vary 
the mass of the lighter scalar $h$ in the range $0.1~{\rm TeV}<m_h<1~{\rm
TeV}$,  
while holding fixed the mass of the heavy scalar $H$ at $m_H=1$ TeV. 

For Model~III, $V V \to t \bar c$ proceeds at tree-level 
via the s-channel neutral Higgs exchange of Fig.~1b.           
Neglecting the SM diagram, the corresponding parton-level cross-section 
${\hat \sigma}_V \equiv {\hat \sigma}(V^1_{h_{V^1}}V^2_{h_{V^2}} \to t \bar c)$
is given by:

\begin{eqnarray}
{\hat \sigma}_V &=
\frac{\left( \sin 2{\tilde \alpha } \right)^2 N_c \pi \alpha^2}{4 \hat s
\beta_V s^4_W}  \left(\frac{m_V}{m_W}\right)^4 
| \epsilon^{V^1}_{h_{V^1}} \cdot \epsilon^{V^2}_{h_{V^2}}|^2  
|\Pi_h - \Pi_H |^2 \times 
\nonumber \\
&  \sqrt { \Delta_t \Delta_c a_+ a_- } ( a_+ \lambda_R^2 + a_- \lambda_I^2 )    
\label{vvhtc}~,
\end{eqnarray}

\noindent where $ \Delta_\ell = m_\ell^2 / \hat s $, 
$ a_\pm = 1 - ( \sqrt {\Delta_t} \pm \sqrt{\Delta_c} )^2$, 
$\beta_\ell \equiv \sqrt {1-4 \Delta_\ell^2}$ and

\begin{equation}
\Pi_{\cal H} = \left( 1 -\Delta_{\cal H}^2 +i \Delta_{\cal H} 
\Delta_{\Gamma_{\cal
H}} \right)^{-1} ~~,~~ \Delta_{\Gamma_{\cal H}}^2 \equiv \Gamma_{\cal H}^2 / \hat s ~. 
\end{equation}

\noindent Given the couplings of Model~III, the width of ${\cal H}$ 
($\Gamma_{\cal H}$) can be readily calculated~\cite{soni3}. The leading 
decay rates in this model are ${\cal H} \to b\bar b ,t \bar t ,ZZ,W^+W^-$ 
and $t \bar c, c \bar t$. We include all these contributions 
when calculating the above cross-sections. In our numerical results we will ignore CP violation and take $\lambda_I=0$ and set $\lambda_R=\lambda$.

Due to the orthogonality properties of the polarization vectors of the two spin one bosons there is no interference between the transverse 
and the longitudinal polarizations. Note that 
$| \epsilon^{V^1}_{\pm} \cdot \epsilon^{V^2}_{\mp}|^2=0$,
$| \epsilon^{V^1}_{\pm} \cdot \epsilon^{V^2}_{\pm}|^2=1$, and $| 
\epsilon^{V^1}_0 \cdot \epsilon^{V^2}_0|^2 = (1+\beta_V^2)^2/(1-\beta_V^2)^2$ 
which grows with $\hat s$. However, we can see from (\ref{vvhtc}), 
that ${\hat {\sigma}}_V(\Delta_{\cal H} \to 0) \to 0$ ensuring unitarity of 
the hard cross-section.    

In general, the transverse distribution functions are bigger than the longitudinal
ones for $x \gsim 0.1$ \cite{wwtt,wapprox1}. Therefore, the relative smallness
of the transverse hard cross-section compared to the longitudinal one is
partly compensated for in the full cross-section. 
In particular, we find that the contribution from the transversely polarized
$W$'s($Z$'s) can give at most $25\%$($35\%$) of the corresponding full
cross-section $\sigma^{\nu \nu tc}$($\sigma^{eetc}$). 

It is evident from (\ref{vvhtc}) that neglecting the small difference between
$m_W$ and $m_Z$ one finds ${\hat \sigma}_W = {\hat \sigma}_Z$. The main difference
between $\sigma^{\nu \nu tc}$ and $\sigma^{eetc}$ arises from the dissimilarity
between the distribution functions for $W$ and $Z$ bosons. In particular,
disregarding the subleading transverse parts of the $WW$ and the $ZZ$ cross-sections,
the relative strength between the $W$ and the $Z$ longitudinal distribution
functions is given by \cite{wwtt,wapprox1}:
  
\begin{equation}
f^Z_0=2c^{-2}_W\left( 2s^4_W - s^2_W + 1/4 \right) f^W_0 \approx
 f^W_0/3 ~, 
\end{equation}

\noindent Therefore, since the cross-sections $\sigma^{\nu \nu tc}$ 
and $\sigma^{eetc}$ are dominated by collisions of longitudinal $W$'s 
and $Z$'s, respectively, $\sigma^{eetc}$ is expected to be smaller by 
about one order of magnitude than $\sigma^{\nu \nu tc}$, which is 
indeed what we find. We will thus only present
numerical results for $\sigma^{\nu \nu tc}$, keeping in mind 
that $\sigma^{eetc}$ exhibits the same behavior though suppressed 
by about one order of magnitude.

Fig.~2 shows the dependence of the scaled cross-section 
$\sigma^{\nu \nu tc}/\lambda^2$ on the mass of the light Higgs $m_h$ for 
four values of $s$ \cite{footscale}. The cross-section peaks at
$m_h \simeq 250$ GeV and drops as the mass of the light Higgs approaches 
that of the heavy Higgs. Indeed as $m_h \to m_H$, 
$\sigma^{\nu \nu tc}/\lambda^2 \to 0$ as expected when $\tilde {\alpha} =\pi/4$ for which the couplings $htc$ and $Htc$ are identical. However, this ``GIM like'' cancelation does not operate when $\tilde {\alpha} \neq \pi/4$ for which $\sigma^{\nu \nu tc}/ 
\lambda^2$ can stay at the fb level even for $m_h \to m_H$ \cite{longerver}. 
When $\sqrt s =2$ TeV the cross-section is about 5 fb for 
$\lambda=1$ and $m_h \approx 250$ GeV\null. Note that the cross-section
scales  
like $\lambda^2$ so that even a moderate change of $\lambda$, say by a factor
of three, can increase or decrease  the cross-section by one order of magnitude.

It is evident from Fig.~2 that, for $m_h\simeq 250$ GeV, in a NLC running
at $\sqrt s \gsim 1$ TeV with an integrated luminosity 
of ${\cal L} \gsim 10^2$ [fb]$^{-1}$, Model~III (with $\lambda=1$) 
predicts hundreds and 
up to thousands of $t \bar c \nu_e {\bar {\nu}}_e$ events,
and tens to hundreds of $t \bar c e^+ e^-$ events.
Note also that even with $m_h \approx 500$ GeV, 
this projected luminosity can  yield hundreds of 
$t \bar c \nu_e {\bar {\nu}}_e$ events and tens of 
$t \bar c e^+ e^-$ events at $\sqrt s=1.5$ TeV\null. The corresponding SM
prediction is of course essentially zero events.

It is also instructive to compare in Model~III the production rate of $e^+e^- \to t 
\bar c \nu_e {\bar {\nu}}_e$ with that of $e^+e^- \to t \bar t \nu_e 
{\bar {\nu}}_e$. We recall that $\sigma^{\nu \nu tt} \equiv 
\sigma(e^+e^- \to W^+ W^- \nu_e {\bar {\nu}}_e \to t \bar t \nu_e 
{\bar {\nu}}_e)$ is dominated by collisions of two longitudinal $W$'s 
at the parton level \cite{wwtt}.
The reaction $W^+W^- \to t \bar t$ can proceed through the $t$-channel
$b$ quark exchange and the s-channel $\gamma,Z,h$ and $H$ exchanges
(the diagrammatic description can be found in \cite{wwtt,ttnunudavid}). 
The helicity amplitudes for $W_L^+W_L^- \to t \bar t$ are:

\begin{eqnarray} 
 &{\cal A}_{\eta = \bar \eta } = \frac{\pi \alpha}{s^2_W}{ m_t \sqrt{ \hat
s } \over  m_W^2} \left( \left[ (1 + \beta_t^2 ) \cos\theta - 2 \beta_t \over
1 + \beta_t^2 - 2\beta_t \cos \theta \right] - 
 \sum_{ {\cal H} = h , H } \beta_t c_{\cal H} b_{\cal H} \Pi_{\cal H} \right)
\!\!\!\!\!\!\!\!\!\!\nonumber\\ 
 &{\cal A }_{ \eta = - \bar \eta } = \frac{2 \pi \alpha}{s^2_W}{ m_t^2 \over
m_W^2} \left( \eta + \beta_t 
\over 1 + \beta_t^2 - 2 \beta_t \cos \theta  \right) \sin \theta
 \label{aetameta}~, 
\end{eqnarray}

\noindent where  $ \eta$ and $\bar \eta $ denote 
the helicities of the $t$ and $ \bar t $ quarks respectively,
$b_{(h;H)} = (-\sin{\tilde {\alpha}} + u \cos{\tilde {\alpha}}; 
\cos{\tilde {\alpha}} + u \sin{\tilde {\alpha}}) $, $ u = \sqrt{2} (
\lambda_R - i \eta \lambda_I / \beta_t ) $;
 $\theta$ is the CM scattering angle, and $c_{\cal H}$ 
is given in (\ref{hww}) \cite{footsmlimit}. For simplicity we
assumed $\lambda_{tt}=\lambda_{tc}=\lambda$. 

In Fig.~3 we plot the ratio $R^{tc/tt} \equiv 
 \sigma^{\nu \nu tc}/\sigma^{\nu \nu tt}$ within Model~III, for $\lambda=1$, ${\tilde {\alpha}}=\pi/4$ and $m_H=1$ TeV as a function of the light Higgs mass, $m_h$, and for $\sqrt s=0.5,1,1.5$ and 2 TeV. We find that  
$\sigma^{\nu \nu tt}$ depends very weakly on $m_h$, with a small 
peak at around $m_h=400$ GeV which fades as $\sqrt s$ grows.
For $\sqrt s=0.5$ TeV and in the range 
$200 ~{\rm GeV} \lsim m_h \lsim 400 ~{\rm GeV}$, $R^{tc/tt}>1$. In particular, for $m_h \approx 250$ GeV, $\sigma^{\nu \nu tc}$ can become 
almost two orders of magnitude larger than $\sigma^{\nu \nu tt}$. 
As $\sqrt s$ grows, $R^{tc/tt}$ drops; in the range 
$200 ~{\rm GeV} \lsim m_h \lsim 400 ~{\rm GeV}$, we find that for 
 $\sqrt s=1$ TeV, $R^{tc/tt}>0.1$, while for $\sqrt s=1.5-2$ TeV, 
$0.01 \lsim R^{tc/tt} \lsim 0.1$.  

The dependence of $\sigma^{\nu \nu tt}$ on $\lambda$ is significant only
near its peak (at $ m_h \sim 400 $ GeV); for 
$200 ~{\rm GeV} \lsim m_h \lsim 400 ~{\rm GeV}$, where $R^{tc/tt}$ 
acquires its largest values, $R^{tc/tt}$ roughly scales as $\lambda^2$. 
Thus, again even a mild change in $\lambda$ can alter $R^{tc/tt}$ appreciably.
Hence, within Model~III, with $m_h$ in
the few-hundred GeV range, it is possible to observe
comparable production rates for the $t \bar c \nu_e {\bar {\nu}}_e$ and 
$t \bar t \nu_e {\bar {\nu}}_e$ even at a NLC running at CM energies of about a TeV.

Next we discuss the two rare decays $t \to W^+W^-c$
and $t \to ZZc$. The latter is, of course, possible only if $m_t > 2m_Z +m_c$. Within the SM these decay channels
are vanishingly small.
For the first one, the tree-level decay, ${\rm Br}(t \to W^+W^-c)
\approx 10^{-13}-10^{-12}$ due
to GIM suppression \cite{jenkins,atwood}. For the second decay the branching
ratio is even smaller since it occurs only at one loop. 

The situation is completely different in Model~III where both
decays occur at the tree-level through the FC Higgs exchange of
Fig.~1b. These decays are related to the fusion reactions
($WW$, $ZZ\to \bar tc$) by crossing symmetry.
Thus in terms of the hard cross-section given in (\ref{vvhtc}): 

\def\qwe{\zeta}

\begin{equation}
\Gamma_V= { m_t^3 \over 32 N_c \pi^2}
\int_{4\qwe_V^2}^{(1- \qwe_c )^2} \!\!\!\!\!\! dz ~
z (z - 4\qwe_V^2)  \!\!\! \!\!\! \sum_{h_{V^1},h_{V^2}} 
\left. {\hat {\sigma}}_V \right|_{\hat s = m_t^2 z}
\end{equation}

\noindent where $\Gamma_V \equiv \Gamma(t \to V V c)$ and 
$\qwe_{\ell} \equiv m_{\ell} /m_t $.  
The scaled-branching-ratio (SBR) $ {\rm Br}(t \to
W^+W^-c)/ \lambda ^2$
is given in Fig.~4; it is largest for $ 2m_W \lsim m_h \lsim m_t$ and drops rapidly in the regions $m_h<2m_W$ or $m_h>200$ GeV. 
For a wide range of $m_h$, the SBR is
many orders of magnitude bigger than the SM. 
Indeed for optimal
values of $m_h$, lying in the very narrow window, $2m_W \lsim m_h \lsim m_t$, the SBR$\sim 10^{-4}$. It is typically a few times $10^{-7}$ for $m_h \gsim m_t$ and can
reach $ \sim 10^{-6}$ in the $m_h \lsim 2m_W$ region.  
Concerning $t \to ZZc$, the branching ratio is typically $\sim 10^{-5}$
for $(2m_Z +m_c) < m_t < 200$ GeV if again $m_h$ lies in a very narrow window, $2m_Z<m_h<m_t$.
Note that, in contrast to the SM, within Model~III, Br($h \to
WW$)$ \sim 1$ for ${\tilde {\alpha}}=\pi/4$; and, even for $ m_h > 2 m_t
$, Br($ h \to WW $)$ \sim 0.7 \gg $Br($ h \to t \bar t$) \cite{notethat}.
Both decays are thus very sensitive to $ m_t $: for $170 ~{\rm GeV}<m_t<200 ~{\rm
GeV}$, a $\sim 15$ GeV shift in $ m_t $ can generate an order of magnitude
change in the Br in the region $2m_V<m_h<m_t$. 

 To summarize, in this paper, we have emphasized the importance 
of searching for the FC reactions, $e^+e^- \to t \bar c \nu_e {\bar {\nu}}_e$ 
and $e^+e^- \to t \bar c e^+ e^-$, in a high energy $e^+e^-$ 
collider. These reactions are sensitive indicators of physics 
beyond the SM with new FC couplings of the top quark. As an 
illustrative example we have considered the consequences of extending 
the scalar sector of the SM with a second scalar doublet such that new 
FC couplings occur at the tree-level. We found that within a large 
portion of the free parameter space of the FC 2HDM, these new FC 
couplings may give rise to appreciable production rates for the 
$t \bar c \nu_e {\bar {\nu}}_e$ and $t \bar c e^+ e^-$ final states 
which can unambiguously indicate the existence of new physics. 
From the experimental point of view, it should be emphasized that 
although $\sigma^{eetc}$ is one order of magnitude smaller 
than $\sigma^{\nu \nu tc}$, the $t \bar c e^+e^-$ signature may be 
easier to detect as it does not have the missing energy associated 
with the two neutrinos in the $t \bar c \nu_e {\bar {\nu}}_e$ final state. 
      
In closing, we also wish to remark that it is most likely that the Higgs particle will have been discovered by the time the NLC starts its first 
run. If indeed such a particle is detected with a mass of a few hundreds GeV, 
it will be extremely important to investigate the 
reactions $e^+e^- \to t \bar c \nu_e {\bar {\nu}}_e$ and 
$e^+e^- \to t \bar c e^+ e^-$ at the NLC as it may serve as a strong 
evidence for the existence of a non-minimal scalar sector with FC 
scalar couplings to fermions. In addition, since supersymmetry strongly 
disfavors $m_h \gsim 200$ GeV, the 
detection of this particle above this limit would encourage the study of a
 general, non-supersymmetric, extended scalar sector.

\bigskip\bigskip

We are grateful to David Atwood, Keisuke Fujii, Daniel Wyler and Marc Sher for discussions.
We acknowledge partial support from U.S. Israel B.S.F. (G.E. and A.S.) and from the U.S. DOE contract numbers DE-AC02-76CH00016(BNL), DE-FG03-94ER40837(UCR). G.E. thanks the Israel Science Foundation and the Fund for the Promotion of Research at the Technion for partial support.

\begin{center}
{\bf Figure Captions}
\end{center}

\begin{description}

\item{Fig. 1:} (a) The Standard Model diagram for $e^+e^- \to t \bar c \nu_e {\bar {\nu}}_e$; (b) Diagrams for $e^+e^- \to t \bar c \nu_e {\bar {\nu}}_e (e^+e^-)$ in Model~III.

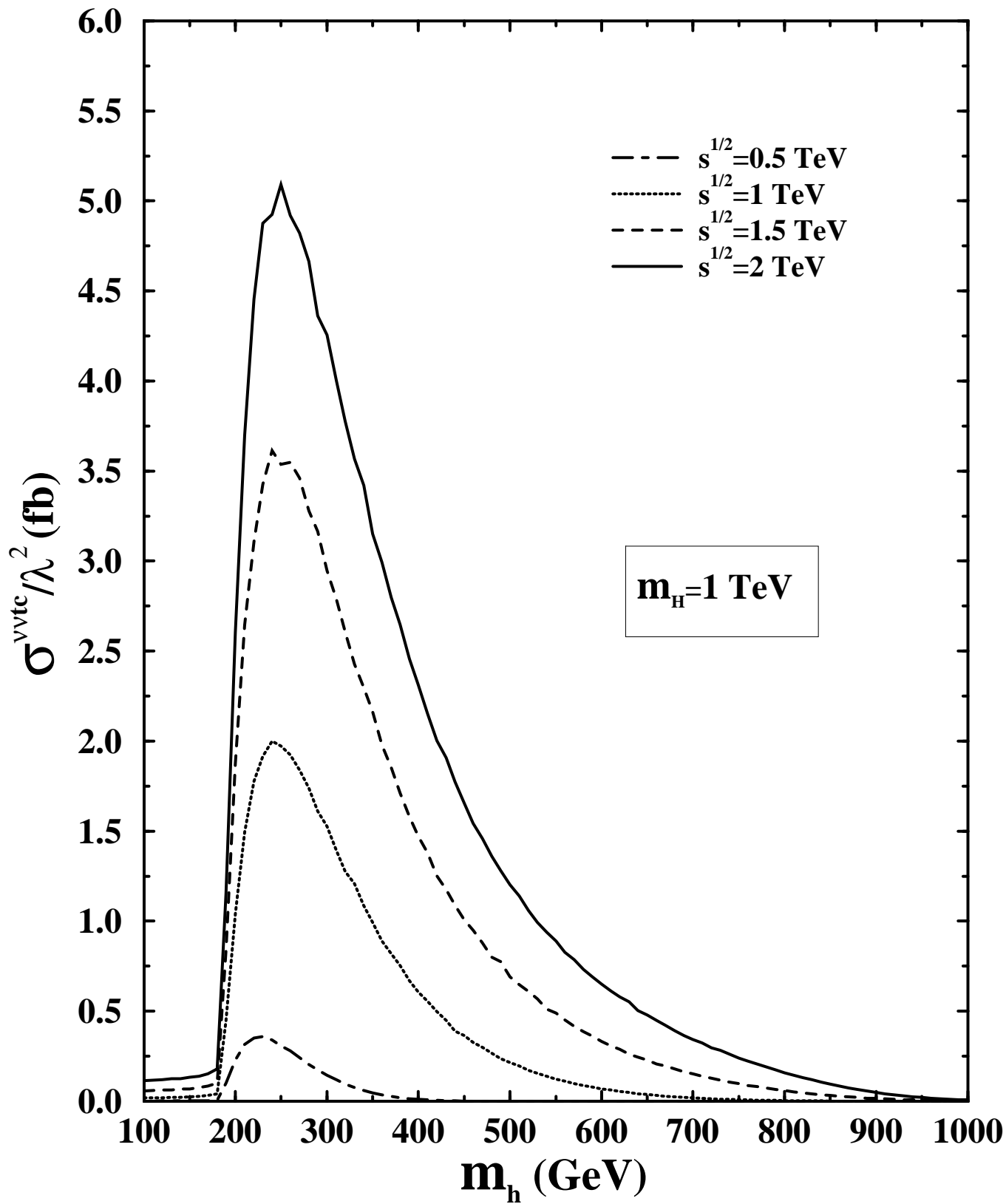
\item{Fig. 2:} The cross-section $\sigma (e^+e^- \to t \bar c \nu_e {\bar {\nu}}_e +  \bar t c \nu_e {\bar {\nu}}_e)$ in units of $\lambda^2$
as a function of $m_h$ for $\sqrt s=0.5,1,1.5$ and 2 TeV. 

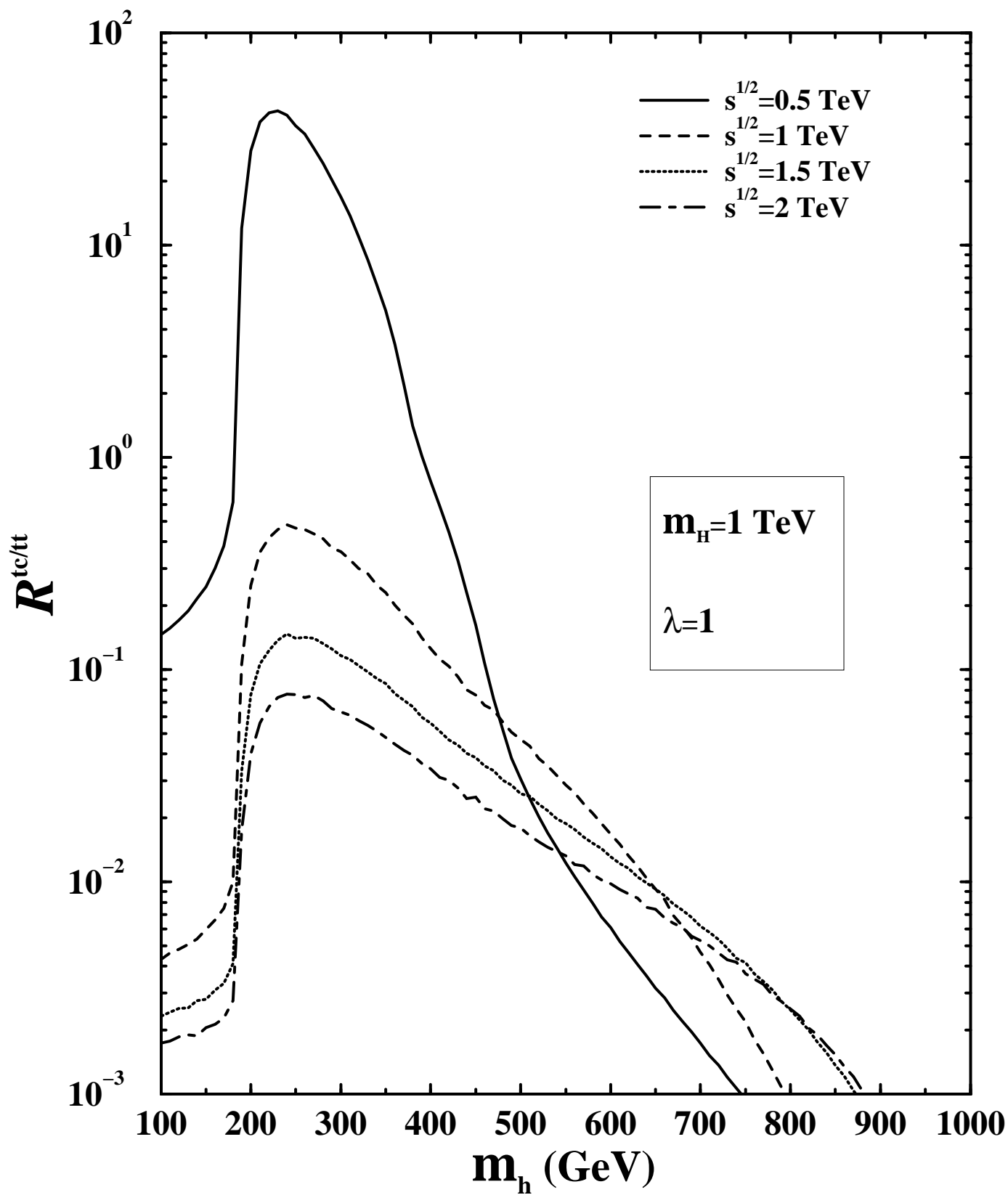
\item{Fig. 3:} The ratio $R^{tc/tt}$ for $\lambda=1$ and $m_H=1$ TeV, as a function of $m_h$ for $\sqrt s=0.5,1,1.5$ and 2 TeV. 

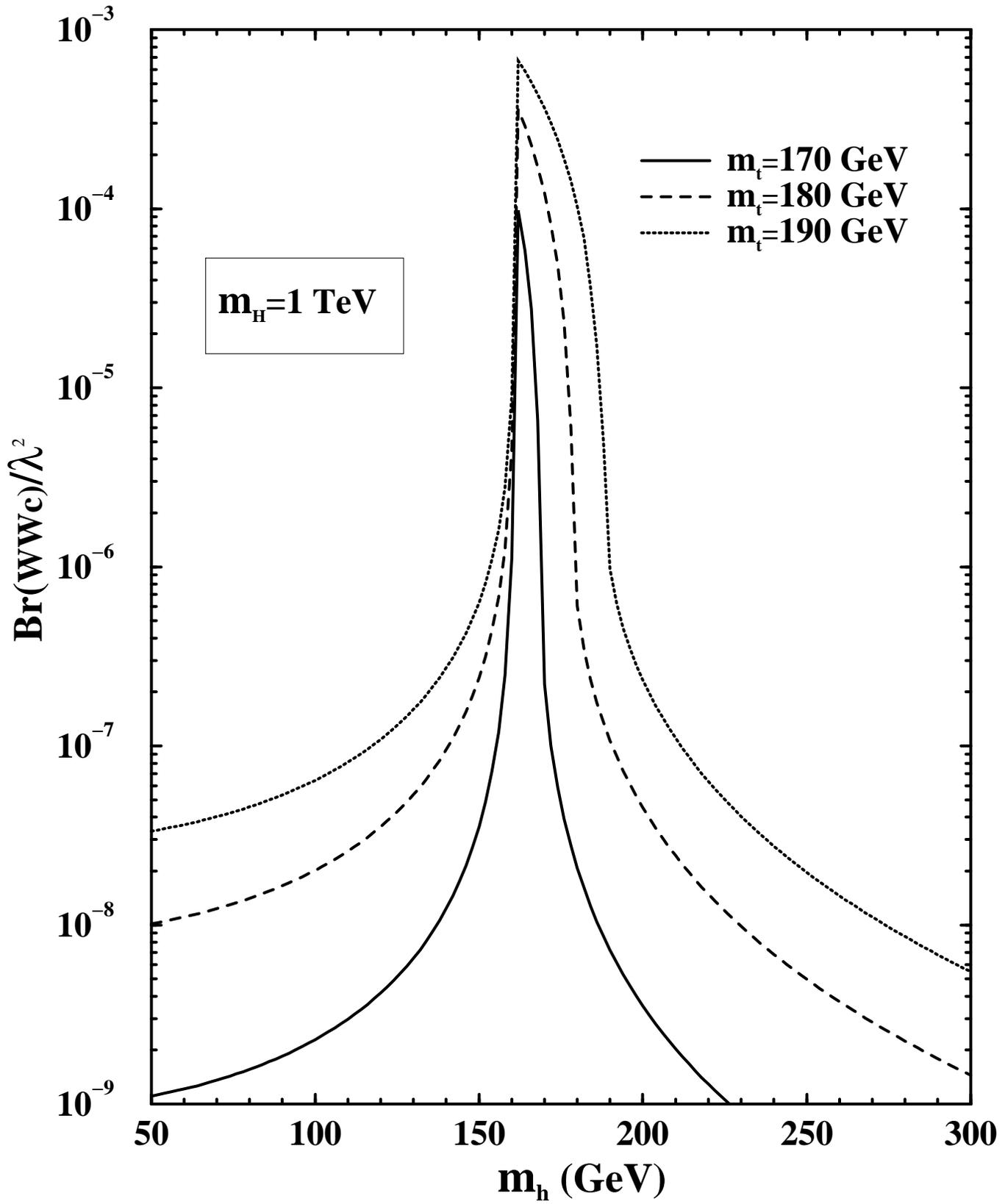
\item{Fig. 4:} The scaled branching ratio, $Br(t\to W^+W^-c)/\lambda^2$
as a function of $m_h$ for various values of $m_t$.

\end{description}

\begin{thebibliography}{99}
\bibitem{lambdatu} $\lambda_{tu}$ is also not well constrained from existing experiments. The Cheng and Sher Ansatz (\ref{csa}) does, of course, imply a much smaller $tu$ coupling compared to the $tc$ one due to the up-charm mass difference.
\bibitem{footvv} We will use the generic notation $V$ for a vector boson i.e., $W^+,W^-$ or $Z$; in most instances the appropriate choice can be fixed by inspection.
\bibitem{footalpha} The cross-section can actually be bigger by about a factor of 3/2 for an optimal choice of ${\tilde {\alpha}} \approx \pi / 6$.
\bibitem{footscale} Note that the scaled cross-section (e.g., $\sigma^{\nu \nu tc}/ \lambda^2$) has a residual mild dependence on $\lambda$ due to the width of the $h$.
\bibitem{notethat} Note that Hou's analytical results
 in Ref.~5 correspond to the choice $\tilde\alpha=0$ in our notation. In this
special case Higgs decays to $WW$, $ZZ$ are suppressed at tree level even when
$m_H>2m_W$. In contrast, for illustrative purposes, we are using
$\tilde\alpha=\pi/4$ in which case $h\to WW$ becomes the dominant
decay.
\end{thebibliography}
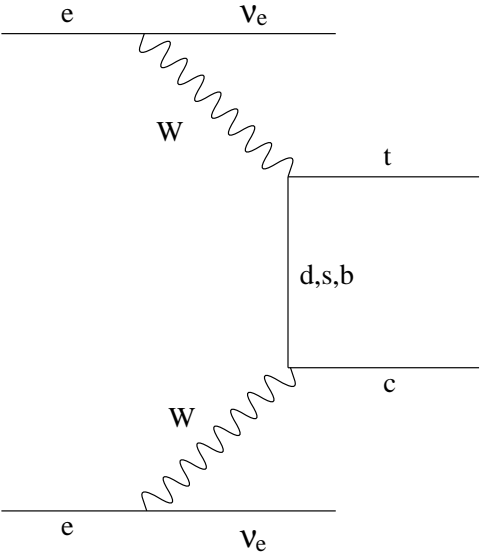
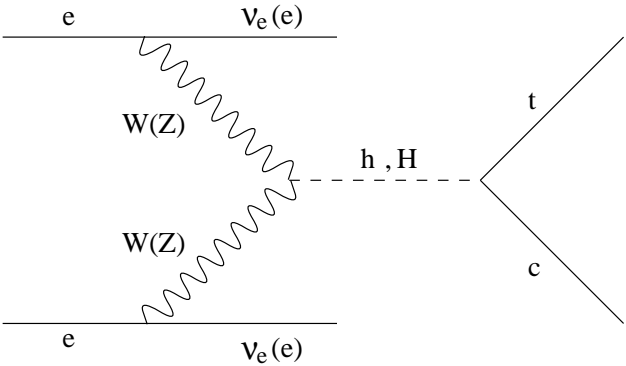
\end{document}